\documentclass[pra,floatfix,twocolumn,showpacs,amsmath,amssymb,final,superscriptaddress]{revtex4}
\usepackage{graphicx} 
\usepackage{dcolumn} 
\usepackage{bm} 
\usepackage[pdftex,colorlinks,bookmarks=false,citecolor=blue,linkcolor=red,urlcolor=blue]{hyperref}
\usepackage{color}

\begin{document}

\title{Fulde-Ferrell-Larkin-Ovchinnikov state in the one-dimensional attractive Hubbard model and its fingerprint in the spatial noise correlations}

\author{Andreas L\"uscher}
\affiliation{Institut Romand de Recherche Num\'erique en Physique des Mat\'eriaux (IRRMA), EPFL, CH-1015 Lausanne, Switzerland}
\author{Reinhard M. Noack}
\affiliation{Fachbereich Physik, Philipps-Universit\"at Marburg, D-35032 Marburg, Germany}
\author{Andreas M. L\"auchli}
\affiliation{Institut Romand de Recherche Num\'erique en Physique des Mat\'eriaux (IRRMA), EPFL, CH-1015 Lausanne, Switzerland}

\date{\today}

\begin{abstract}
We explore the pairing properties of the one-dimensional attractive Hubbard model in the presence of finite spin polarization. 
The correlation exponents for the most important fluctuations are determined as a function of the density and the polarization. 
We find that in a system with spin population imbalance, Fulde-Ferrell-Larkin-Ovchinnikov (FFLO)-type pairing at wavevector 
$Q=|k_{F,\uparrow}-k_{F,\downarrow}|$ is always dominant and there is no Chandrasekhar-Clogston limit. 
We then investigate the case of weakly coupled 1D systems and determine the region of stability of the 1D FFLO phase.
This picture is corroborated by density-matrix-renormalization-group (DMRG) simulations of the spatial noise correlations in uniform and trapped systems,
unambiguously revealing the presence of fermion pairs with nonzero momentum $Q$. 
This opens up an interesting possibility for experimental studies of FFLO states.
\end{abstract}


\maketitle

\section{Introduction}
In two-component Fermi systems with attractive interactions, in which
both species share the same Fermi surface, pair formation is
successfully described by the Bardeen-Cooper-Schrieffer (BCS)
theory~\cite{bardeen57}. This scenario is, for instance, realized in
conventional superconductors, where electrons of opposite spin and
momentum form Cooper pairs. In contrast, if the Fermi surfaces are
different, the situation is subtler because
not all of the fermions can form pairs. The question thus arises as to
whether a superfluid phase exists in this setup and if so, what its
properties are.
On the theoretical side, different answers to this question have been
proposed~\cite{sarma63,fulde64,larkin64,muther02,liu03,bedaque03,yi06}
ranging from exotic superfluid phases to more conventional phase
separations. Particular attention has been devoted to the so-called
Fulde-Ferrell-Larkin-Ovchinnikov (FFLO, sometimes also denoted LOFF in
the literature) state, in which fermion pairs with nonzero momentum form an inhomogeneous superfluid phase. Initially developed for superconductors in magnetic fields, this theory has been applied to heavy fermion systems~\cite{radovan03,bianchi03} and dense quark matter~\cite{casalbuoni04}.

\begin{table*}
\centering
\caption{\label{tab:correlations}
\emph{Correspondence between the correlation functions of the attractive and the repulsive Hubbard model. Note that the mapping holds for all three types of the triplet superfluidity.  From Ref.~\onlinecite{singh91}.}} 
\vspace{1mm}
\begin{ruledtabular}
\begin{tabular}{l | c||c | l}
Attractive interaction & \hspace*{1cm} $U<0$ \hspace*{1cm} & \hspace*{1cm}   $U>0$\hspace*{1cm} & Repulsive interaction \\[1mm] \hline &&&\\[-2mm]
density correlations (CDW) & $n_l n_0$ & ${\bar S}_l^z {\bar S}_0^z$ & longitudinal spin correlations (lSDW) \\[1mm]
singlet superfluidity (SS) & $c_{l\uparrow}^\dag c_{l\downarrow}^\dag c_{0\uparrow}^{\phantom{\dag}} c_{0\downarrow}^{\phantom{\dag}}$ & $\left(-1\right)^l {\bar S}_l^- {\bar S}_0^+$ & transverse spin correlations (tSDW)\\[1mm]
triplet superfluidity (TS) & $c_{l\uparrow}^\dag c_{l+1,\uparrow}^\dag c_{0\uparrow}^{\phantom{\dag}} c_{1\uparrow}^{\phantom{\dag}}$ & ${\bar c}_{l\uparrow}^\dag {\bar c}_{l+1,\uparrow}^\dag {\bar c}_{0\uparrow}^{\phantom{\dag}} {\bar c}_{1\uparrow}^{\phantom{\dag}}$ & triplet superfluidity (TS) \\[1mm]
longitudinal spin correlations (lSDW) & $S_l^z S_0^z$ & ${\bar n}_l {\bar n}_0$ & density correlations (CDW) \\[1mm]
transverse spin correlations (tSDW) & $\left(-1\right)^l S_l^- S_0^+$ & ${\bar c}_{l\uparrow}^\dag {\bar c}_{l\downarrow}^\dag {\bar c}_{0\uparrow}^{\phantom{\dag}} {\bar c}_{0\downarrow}^{\phantom{\dag}}$ & singlet superfluidity (SS)
\end{tabular}
\end{ruledtabular} 
\end{table*}

Recent advances in methods for trapping and controlling ultracold
atoms have opened up the possibility of experimentally
observing population imbalanced mixtures of ultracold
fermions~\cite{zwierlein06,partridge06,zwierlein06c,shin06,schnuck07}.
Compared to solid state materials, cold gases allow an unprecedented
control over the Fermi surface mismatch by adjusting the population
imbalance
\begin{equation*}
p=\frac{N_\uparrow-N_\downarrow}{N_\uparrow+N_\downarrow} \ .
\end{equation*}
Here $N_\sigma$ denotes the number of particles in one of two hyperfine states
labeled by a pseudo-spin index
$\sigma \in \left\{\uparrow,\downarrow\right\}$. So far, these
experiments have not revealed any of the proposed exotic superfluids,
but have
concluded that the system either phase separates, forming
an unpaired normal state immersed in a BCS
superfluid~\cite{partridge06}, or forms pairs that do not
condense~\cite{schnuck07}. Although theoretical studies of the
stability and extent
of the FFLO state in the phase diagram have not yet reached a
consensus, it is clear by now that it is best realized at low
temperatures and in quasi one-dimensional (1D)
geometries~\cite{mizushima05,orso07,hu07,parish07}. Given the disagreement between
theoretical predictions and the difficulty of unambiguously
identifying exotic phases encountered in experiments, we believe it 
lends insight
to study the simplest possible system that exhibits FFLO-type
pairing: the 1D Hubbard model with attractive interactions. On the one
hand, this system is well suited for theoretical studies
because it is exactly solvable via the Bethe-Ansatz and is numerically exactly
treatable in density-matrix-renormalization-group (DMRG) simulations. 
On the other hand, it has now become possible to realize it
experimentally~\cite{moritz05}. 
This paper consists of two parts:
In the first, we derive various correlation exponents as a function of the density and the polarization 
and demonstrate that the superfluid FFLO correlations
dominate at all polarizations and that there is no
Chandrasekhar-Clogston limit~\cite{chandrasekhar62,clogston62} in the attractive 1D Hubbard model. 
We also address the stability of the FFLO phase in weakly coupled chains.
In the second part, we calculate the experimentally accessible spatial noise
correlations quantitatively for uniform and trapped systems and show that FFLO states leave a
distinct fingerprint, making it an ideal tool to detect and
characterize exotic pairing phases~\cite{altman04,yang05}. 

\section{The model}
We concern ourselves
with a population-imbalanced 1D Fermi gas with attractive interactions,
which can be described by the Hubbard model
\begin{equation} \label{eq:H}
H=-t \sum_{l,\sigma} \left( c_{l,\sigma}^\dag c^{\phantom{\dag}}_{l+1,\sigma} + \text{H.c.}\right)
+ U \sum_{l} n_{l,\uparrow} n_{l,\downarrow} \ .
\end{equation}
Here $c_{l,\sigma}^\dag$ ($c^{\phantom{\dag}}_{l,\sigma}$) creates
(destroys) a fermion with pseudo-spin $\sigma$ on site $l$ and
$n_{l,\sigma}= c_{l,\sigma}^\dag c^{\phantom{\dag}}_{l,\sigma}$ is the
occupation number operator. The parameters $t$ and $U < 0$
characterize the nearest-neighbor hopping and the attractive on-site
interaction, respectively. 
Furthermore, we specify the polarization $p$ and the particle density
$n=(N_{\uparrow}+N_{\downarrow})/L$, $L$ being the length of the
chain. The physics of the attractive Hubbard chain for zero
polarization is quite transparent: the on-site attraction $U$ favors
locally doubly occupied sites. These doubly occupied sites can lead to
$2 k_F$ charge-density wave (CDW) fluctuations and to singlet
superfluid (SS) fluctuations of pairs with zero momentum. Away
from half filling, the pairing fluctuations dominate for all densities
when $U<0$. Since spin excitations are gapped in this phase, spin density
wave (SDW) and triplet superfluid (TS) correlations decay
exponentially for $U<0$. The correlation functions are defined in Table~\ref{tab:correlations}.
As soon as the system is polarized, the spin gap closes
and all the above-mentioned correlation functions decay as power
laws. Due to the unequal number of up and down particles, the corresponding
Fermi vectors 
$k_{F\sigma} = N_\sigma\pi/L = n\left(1\pm p\right)\pi/2
$
are different, and one might expect that the singlet superfluid
correlations still remain important, albeit now composed of pairs with
finite momentum $Q=|k_{F\uparrow}-k_{F\downarrow}|=n p \pi$. 
This is the prediction of a bosonization analysis~\cite{yang01}.

\section{Correlation exponents for $p>0$}
It is well known~\cite{emery76,moreo07} that the attractive Hubbard model can be mapped onto the repulsive one under a staggered particle-hole transformation of one of the two species. To be specific, let us use the transformation 
\begin{align*}
c^\dagger_{l \uparrow} &= (-1)^l {\bar c}^{\phantom{\dagger}}_{l\uparrow}  \ , & 
c^\dagger_{l \downarrow} &={\bar c}^\dagger_{l \downarrow} \ .
\end{align*}
The sector with particle density $n$ and polarization $p$ of the attractive model maps onto 
${\bar n}=1-np$ and ${\bar p}=(1-n)/(1-np)$
on the repulsive side 
(symbols with a bar denote quantities for the repulsive model). 
Under this transformation, density correlations in the attractive model map onto longitudinal spin correlations in the repulsive model, while singlet superfluidity is mapped onto transverse spin correlations, see Table~\ref{tab:correlations}. The Fermi momentum of the transformed species is shifted, ${\bar k}_{F,\uparrow}=\pi-k_{F,\uparrow}$, while ${\bar k}_{F,\downarrow}=k_{F,\downarrow}$ is left unchanged. 

\begin{figure*}
\centerline{\includegraphics[width=0.95\linewidth,clip]{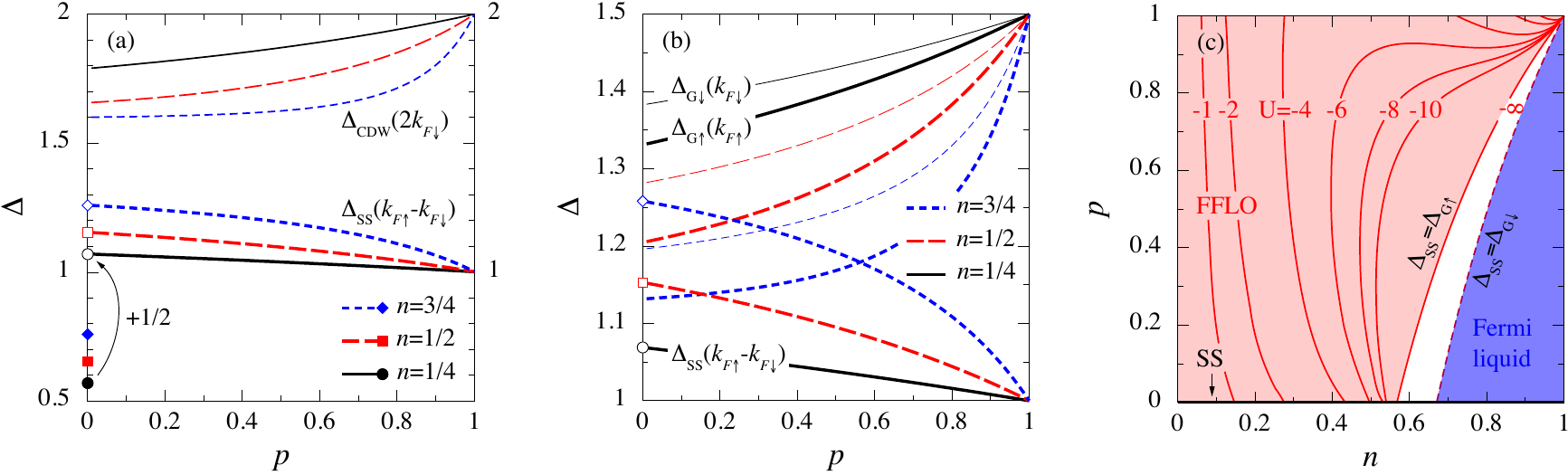}}
\caption{
\emph{(Color online). Correlation exponents of the polarized attractive Hubbard model. (a)
Dominant two-particle correlation exponents $\Delta_\xi\left(k\right)$ in the strong-coupling limit  $(|U|\gg t)$ as a function of polarization $p$ for different densities
$n$. Singlet superfluid fluctuations (SS) at momentum
$|k_{F,\uparrow}-k_{F,\downarrow}|$ are dominant, while
$2k_{F\downarrow}$ charge density waves (CDW) are the subleading
fluctuations. Note that the exponents  $\Delta_\mathrm{SS}$ differ by $1/2$ for $p=0$ and
$p\to0^+$, as predicted in Ref.~\onlinecite{yang01}. (b) Evolution of the Green's function correlation exponents with polarization for different densities in the strong-coupling limit. The SS exponents are shown again for comparison. The Green's function of the majority particles ($\Delta_{\mathrm{G}\uparrow}$) always decays more slowly than the minority one ($\Delta_{\mathrm{G}\downarrow}$). (c) Phase diagram for an infinite number of weakly coupled chains as a function of density and polarization for various interactions $U$. In the FFLO phase, SS correlations are dominant, while in the Fermi liquid phase, the asymptotic decay of the Green's functions is slower than that of any two-particle correlator. For finite $U$, only the boundaries determined by the majority particles ($\Delta_{\mathrm{SS}}=\Delta_{\mathrm{G}\uparrow}$) are shown. SS denotes the conventional balanced $s$-wave superfluid for all $n$ at $p=0$.} 
\label{fig:exponents}}
\end{figure*}

With this correspondence in hand, 
we can use the results of Frahm and Korepin~\cite{frahm} on the correlation exponents for the repulsive Hubbard model in a magnetic field and translate the behavior of the correlation functions to the attractive case. In general, the leading asymptotic behavior of an equal-time correlator can be written as 
\begin{equation} \label{eq:correlations}
C_\xi(l)=
A_\xi \frac{\cos \left(\alpha_\xi k_{F\uparrow} l +\beta_\xi k_{F\downarrow} l\right)}{\left| l \right|^{\Delta_\xi\left(\alpha,\beta\right)}} \ ,
\end{equation}
where the subscript $\xi$ denotes the type of fluctuation, see Table~\ref{tab:correlations}.
The correlation exponents $\Delta_\xi$ can be derived from the elements of the dressed charge matrix, while the coefficients $\alpha$ and $\beta$ are constrained by the parity of the quantum numbers of the intermediate states~\cite{frahm}.
Note that  Eq.~(\ref{eq:correlations}) only gives the leading term of the asymptotic behavior of the correlation functions. In principle there are also higher harmonics present. However, their correlation exponents being larger, one can safely neglect these contributions if one is interested in the asymptotic behavior. Our numerical results presented in the next section show that the dominant FFLO mode occurs at the expected single momentum $Q$.
A numerical solution of the Bethe-Ansatz equations for the Hubbard model yields the elements of the dressed charge matrix, which allow one to sketch the the evolution of the correlation exponents with density and polarization~\cite{frahm}.
Our results for the dominant two-particle correlators obtained in the strong-coupling limit $U\gg t$ are shown in Fig.~\ref{fig:exponents}(a).
First, for $p=0$, we plot the values of
$\Delta_\mathrm{SS}(k=0)$ for three different densities (filled symbols). In this limit, the CDW correlation exponent is given by
$\Delta_\mathrm{CDW}(2k_F) = 1/\Delta_\mathrm{SS}(0)>1$. Second, for
$p\rightarrow0^+$ it has been predicted by Yang~\cite{yang01} that the
FFLO  $\Delta_\mathrm{SS}\left(k_{F\uparrow}-k_{F\downarrow}\right)$
exponent is given by the SS exponent at $p=0$ plus 1/2 (open symbols). Our
data confirm this result, see also Ref.~\onlinecite{frahm08}. 
Then, upon increasing the population imbalance, this correlation
exponent
decreases monotonically until it reaches unity as $p\rightarrow 1$. 
The leading contribution to the asymptotic behavior of the CDW
correlations is found for wave vectors $2k_{F\downarrow}$, but their
correlation exponent is always sizably larger than the FFLO exponent
and increases towards $2$ as $p\rightarrow 1$. 
Note that even in the limit $p \rightarrow 0^+$ the density correlation exponents 
at $2k_{F\downarrow}$ and $2k_{F\uparrow}$ do not become equal, signaling the 
absence of spin-charge separation at all $p>0$ in the general case $n\neq 1$~\cite{rizzi07, frahm08}.

Although longitudinal
SDW and TS correlations also decay as power laws, they are not shown
here because they decay 
more rapidly
(with exponents $\Delta>2$). 
These results clearly show that FFLO pairing is the
dominant two-particle correlation for any nonzero polarization; hence, there is no
Chandrasekhar-Clogston limit~\cite{clogston62,chandrasekhar62} beyond
which the pairing breaks down in this strictly 1D model.

While the above analysis of the two-particle correlation exponents shows that pairing fluctuations dominate, the 1D character of the system prevents the formation of a truly long-range-ordered SS phase. In higher dimensions, the restriction to power-law decay is lifted and one expects that by coupling an infinite number of chains with a suitable interchain interaction, a long-range ordered FFLO phase can be stabilized. Following the standard analysis of the dimensional crossover, see, e.g., Ref.~\onlinecite{giamarchi03}, we compare the dominant two-particle correlations with the Green's functions and argue that upon lowering the temperature of the system, one either enters a phase dominated by the physics of 1D, if the two-particle correlations decay more slowly than the Green's function, or, one flows first to a higher dimensional Fermi liquid if the opposite condition is satisfied. 

In Fig.~\ref{fig:exponents}(b), we plot the dependence of the SS and the two Green's function correlation exponents on the polarization $p$ for different fillings $n$ in the strong-coupling limit $U\gg t$. In the unpolarized system, the Green's function decays exponentially because of the spin gap. In the polarized regime, the exponent of the majority particles is smaller than the minority one, but both tend to $3/2$ in the limit $p \to 1^-$. Fig.~\ref{fig:exponents}(c) shows the phase diagram obtained by comparison of the single- and two-particle correlation exponents as a function of density and polarization for various interactions $U$. Here we go beyond Ref.~\onlinecite{yang01}, which neglected the fact that spin and charge are coupled immediately for all $p>0$ sufficiently away from half filling $n=1$. 
Because of the difference between the majority and minority exponents, the phase boundary between the FFLO and the Fermi liquid regime is smeared out, i.e., there are two boundaries determined by $\Delta_\mathrm{SS}=\Delta_{\mathrm{G}\uparrow}$ and $\Delta_\mathrm{SS}=\Delta_{\mathrm{G}\downarrow}$. For simplicity, the two boundaries are only shown for the strong-coupling limit, while for finite $U$, only the crossover with the majority particles ($\Delta_\mathrm{SS}=\Delta_{\mathrm{G}\uparrow}$) is indicated.
In dilute systems, FFLO pairing is observed for all polarizations, while at higher fillings, the superfluid phase is only established in situations with strong population imbalance. This leads to two disconnected patches of FFLO ordering for interactions $U\lesssim4$, as can be seen in Fig.~\ref{fig:exponents}(c). For instance, for $U=-1$, the FFLO phase is realized in the region $n\lesssim0.1$, but there is also a small reentrant patch in the vicinity of the fully polarized half-filled system at $n\approx1$ and $p\approx 1$. 
Note that this simple comparison of the correlation exponents does not take into account residual interactions that might destabilize the Fermi liquid phase at low temperatures and neglects the precise nature of the interchain couplings. 
For a parallel discussion of the dimensional crossover, we refer the reader to a very recent work by Zhao and Liu~\cite{zhao08}, in which a similar phase diagram has been presented.

\begin{figure*}
\includegraphics[width=0.75\textwidth,clip]{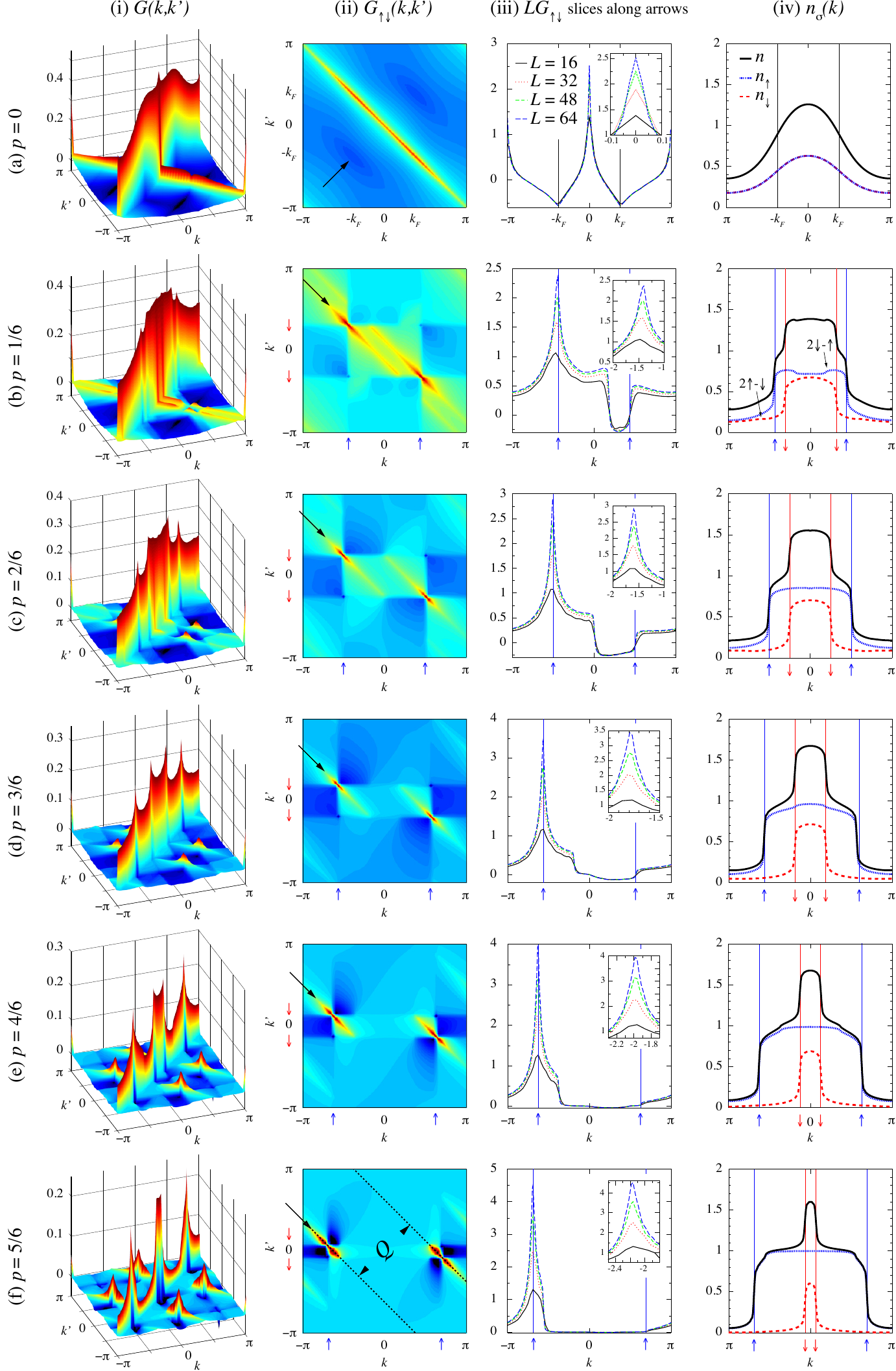}
\caption{
\emph{
  (Color online). Noise correlations in the attractive 1D Hubbard
  model for several spin polarizations $p$ at density $n=3/4$. In the
  unpolarized case [row (a)], one identifies dominating $s$-wave
  superfluid correlations in $G_{\uparrow\downarrow}$, (a,ii), along
  the anti-diagonal, indicating the presence of pairs with zero total
  momentum. In a population-imbalanced system [rows (b) to (f)], the
  FFLO state is clearly visible in $G_{\uparrow\downarrow}$ [column (ii)] and
  also $G$ [column (i)] as strongly peaked signals that shift apart 
  as
  the polarization increases. The separation $Q$ [indicated in (f,ii)]
  between the peaks is equal to the total momentum of the
  pairs. Finite-size-scaling extrapolations of $L G_{\uparrow\downarrow}$ taken along
  $q'=-q-k_{F\uparrow}+k_{F\downarrow}$, indicated by the arrows shown
  in column (iii), illustrate the growth of the signal with increasing
  system size. The momentum distributions [column (iv)] show a clear
  difference between the gapped ($p=0$) and the gapless ($p>0$) phases,
  but do not contain
  direct evidence for the presence of
  an FFLO state.
} 
\label{fig:noise}}
\end{figure*}
\section{Noise correlations}
Recently, Altman {\it et al.}~\cite{altman04} suggested using spatial noise correlations as a universal probe of fluctuations in cold atomic gases. In experiments, these correlations can be extracted from a statistical
analysis of many time-of-flight images of the expanding cloud of
atoms, once they have been released from the trap. A key ingredient of
the method is the fact that the spatial density distribution of the
expanding cloud becomes proportional to the momentum distribution in
the interacting system after the trap is released. For the lattice model we consider in this work, spatial noise correlations are defined as
\begin{equation} \label{eq:G}
G_{\sigma\sigma'}\left(k,k'\right) = 
\left\langle n_{k,\sigma} n_{k',\sigma'}\right\rangle -
 \left\langle n_{k,\sigma}\right\rangle \left\langle n_{k',\sigma'}\right\rangle \ ,
\end{equation}
where $n_{k,\sigma}=c_{k,\sigma}^\dag c_{k,\sigma}^{\phantom{\dag}}$ is the occupation number operator in momentum space. 
We will also refer to the total noise correlation 
\begin{equation*}
G\left(k,k'\right) = \sum_{\sigma,\sigma'} \ G_{\sigma\sigma'}\left(k,k'\right) \ .
\end{equation*}

\subsection{Strong-coupling limit}
We now proceed to a quantitative numerical evaluation of the noise correlations in the attractive Hubbard model with strong interaction $U/t=-10$, at filling $n=3/4$ for different polarizations $p$, using a coordinate-space-based DMRG algorithm with open boundary conditions~\cite{white92,schollwoeck05}. For an in-depth discussion of the method and the noise correlations of various phases of extended Hubbard models with balanced spin populations, we refer the reader to Ref.~\onlinecite{luscher07}. To determine the noise correlations of a system with $L=64$ sites, one has to calculate $4 L^4\approx 67 \cdot 10^6$ four-point correlators. This number could be reduced by exploiting symmetries; however, we choose not to do so, to get unbiased results for the sum rules~\cite{luscher07}, which we use to check convergence. While a system with 64 sites seems rather small for DMRG calculations, these sizes are, however, sufficient for the purpose of the present work. The main features of the noise correlations are already well established in these systems and finite-size scalings do not show any ambiguities. For such moderate system sizes, keeping up to 400 states results in a discarded weight smaller than $10^{-7}$.

Let us start the analysis with the unpolarized system, whose shot
noise is shown in Fig.~\ref{fig:noise}[row (a)]. In this case,
fermions tend to pair in on-site singlets, leading to a gap in the
spin sector and thus exponentially decaying SDW and TS
correlations. 
Since the correlation exponents $\Delta_{SS} <\Delta_{CDW} $, 
the SS correlations are dominant over the CDW 
correlations.
This can be seen in the shot noise. Quite generally,
particle-hole fluctuations appear as negative signals along $k'=k\pm
q$, where $q$ is the ordering wave vector, while particle-particle
correlations give a positive contribution along $k'=-k\pm q$~\cite{altman04,mathey05,luscher07}. The
dominant positive signal of the total shot noise G around 
$\left(\pm k_F,\mp k_F\right)$, shown in Fig.~\ref{fig:noise}(a,i), thus clearly
indicates that SS correlations dominate over CDW fluctuations.

As soon as the system is polarized, one
recognizes two important modifications of the noise correlations with
respect to the unpolarized case, see 
Figs.~\ref{fig:noise}[rows (b) to (f)].
Most noticeably,
the distinct comb present in the $\uparrow\downarrow$-channel splits
into two ridges peaked at $\left(\pm k_{F\uparrow},\mp k_{F\downarrow}\right)$, which move away from the anti-diagonal with
increasing imbalance. The splitting
$Q=|k_{F\uparrow}-k_{F\downarrow}|$ between them directly corresponds
to the momentum of the Cooper pairs. 
{\em This is the fingerprint of the FFLO state.} The abrupt change from a comb
with roughly constant height to pronounced peaks is due to the sudden increase 
of the correlation exponent upon entering the polarized phase. 
\begin{figure}
\includegraphics[width=0.4\textwidth,clip]{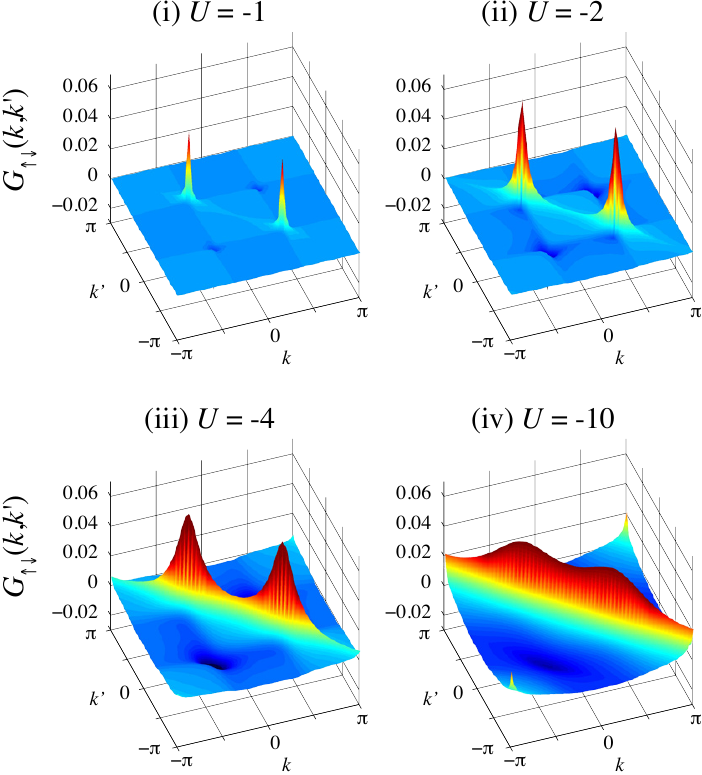}
\caption{
\emph{(Color online).
Off-diagonal noise correlations $G_{\uparrow\downarrow}$ in an unpolarized system with $L=64$ sites at filling $n=3/4$ for different interactions $U$. The Fermi surfaces (nine-tile pattern), visible for small interactions, become weaker and finally disappear for strong couplings. SS correlations develop around opposite Fermi points and start to spread out along the whole anti-diagonal with increasing attraction.
} 
\label{fig:variableU}}
\end{figure}
In the same spirit, we
note that the peaks become more pronounced for increasing $p$, in agreement
with the fact that the FFLO exponent decreases with increasing $p$,
see Fig.~\ref{fig:exponents}(a). 
Secondly, traces of the Fermi surface, which are completely
absent in the unpolarized system due to the presence of a spin gap, appear
in all parts of the noise correlations as a nine-tile pattern best visible in Figs.~\ref{fig:noise}[column (ii)]. 
These Fermi surfaces are characteristic of a system with gapless excitations~\cite{luscher07}.
Based on the fact that particle-hole correlations give rise to a negative signal in the noise correlations while particle-particle contributions are 
positive, we expect four of the nine tiles to have negative values. This is in agreement with our numerical results.
A density plot, 
along with plots of slices
along the diagonal of $G_{\uparrow\downarrow}$, allows one to study the
FFLO phase in more detail. First, as derived before, the peaks of the
SS signal in the $\uparrow\downarrow$-channel are located at
$\left(\pm k_{F\uparrow}, \mp k_{F\downarrow}\right)$ and the comb
extends along $k'=-k\pm\left(k_{F\uparrow}-k_{F\downarrow}\right)$.  
In the finite-size scalings of the SS signals parallel to the anti-diagonal, see Figs.~\ref{fig:noise}[column (iii)], one can see that the correlations are peaked at $\left(\pm k_{F\uparrow}, \mp k_{F\downarrow}\right)$ and drop down to a constant positive background rather quickly away from this point. This phenomenon reflects the fact that the pairing of particles is stronger the closer the two momenta are to the corresponding Fermi vectors. Going along the direction of one of the arrows in the density plots, see Figs.~\ref{fig:noise}[column (ii)], one encounters a first peak representing the pairing of down particles at momentum $k'\approx k_{F\downarrow}$ with up particles at momentum $k\approx -k_{F \uparrow}$. Continuing along this line, the $k'$ momentum crosses the horizontal axis and approaches the Fermi point $-k_{F\downarrow}$. Although pairing in this region is kinematically possible, it is clear that down particles around $k'\approx -k_{F\downarrow}$ are preferably matched with up particles from the other side of the Brillouin zone at $k_{F\uparrow}$. This illustrates the fact that the pairing correlations are closer to a Larkin-Ovchinnikov~\cite{larkin64} type involving both $\pm Q$ pairs than a Fulde-Ferrell~\cite{fulde64} type with a single $Q$.  Concerning experiments, it is encouraging to see that state-selective measurements of
the shot noise, i.e., separate measurements of the different
$G_{\sigma \sigma'}$ are not necessary 
in order to identify the FFLO phase, since at least in the highly polarized regime, the superfluid features present in the total shot noise $G$ are well separated and almost as pronounced as the main signal along the diagonal, see Figs.~\ref{fig:noise}[bottom of column (i)]. 

The occupation number $n_\sigma\left(k\right)$ for the two species
separately as well as their sum is shown in Figs.~\ref{fig:noise}[column (iv)].
The main characteristics for all nonzero polarizations is
the sharp transition from empty to occupied orbitals and vice-versa at
momenta $\left| k_{F\sigma}\right|$, similar to the abrupt change
observed in the momentum distribution of a metallic Tomonaga-Luttinger liquid, see Fig.~6(d)
in Ref.~\onlinecite{luscher07}. In contrast, in the spin-gapped phase
at zero polarization, the Fermi points are invisible. We note that 
these density profiles do not
contain direct evidence for the presence of dominant FFLO
correlations, since, for example, the population-imbalanced repulsive
Hubbard chain would show a very similar $n_\sigma\left(k\right)$
without having dominant FFLO correlations.
\begin{figure*}
\includegraphics[width=0.75\textwidth,clip]{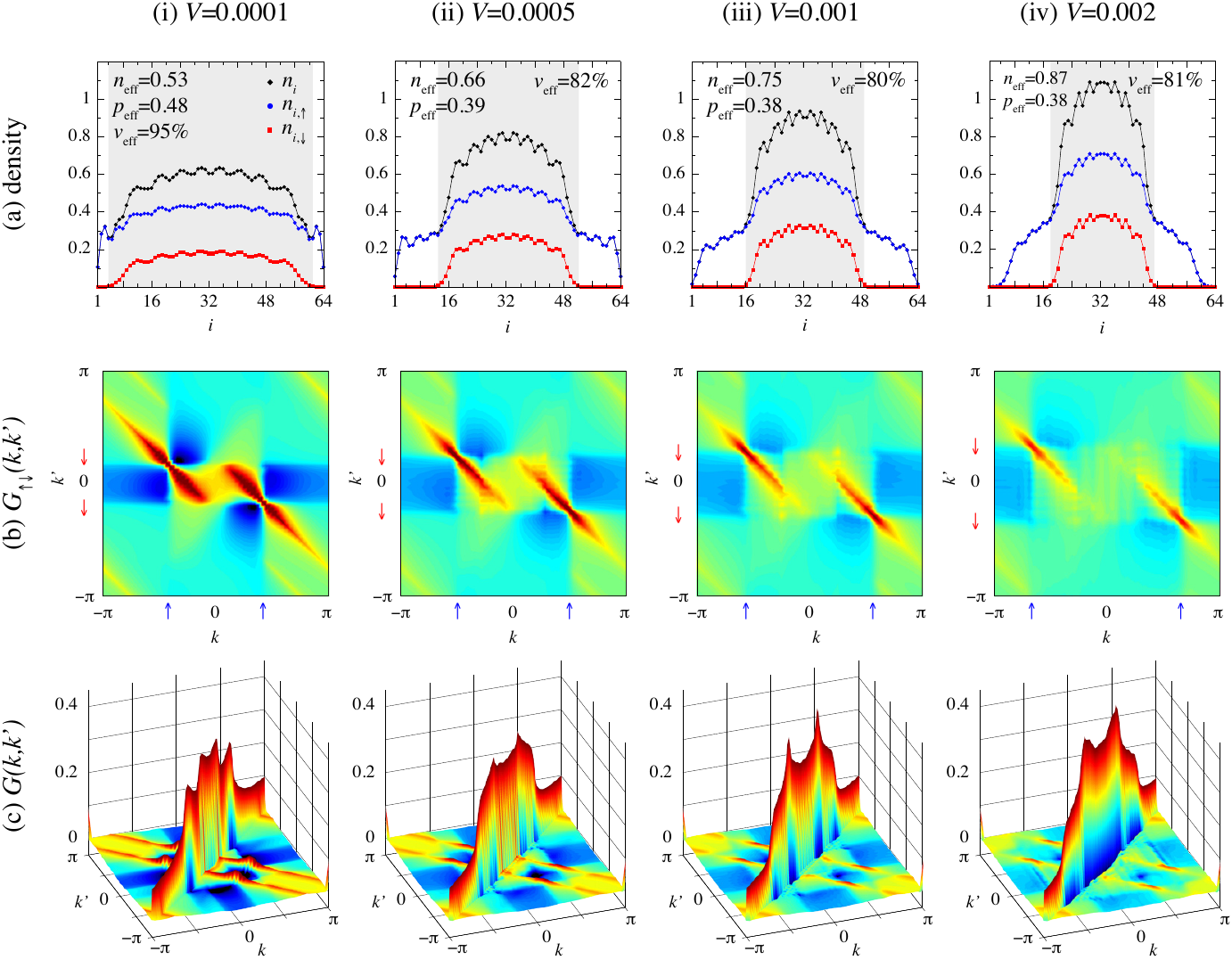}
\caption{
\emph{
  (Color online). Local densities for systems with $L=64$ sites in the strong-coupling limit $U/t=-10$ at filling $n=1/2$ and polarization $p=1/2$ 
for various depths of the trap V [row (a)]. Even for small confinement, some majority particles are pushed away from the center and
 are arranged in fully polarized wings. In the central area of the trap (shaded area), an FFLO state at an effective filling $n_\text{eff}$ and polarization
 $p_\text{eff}$ is formed, as can be deduced from the characteristic signal in the off-diagonal noise correlator $G_{\uparrow\downarrow}$ [row (b)] and in the total shot noise $G$ [row (c)].
} 
\label{fig:trap}}
\end{figure*}

\subsection{Away from strong-coupling}
So far, we have focused on the strong-coupling regime, taking $U/t=-10$. Reducing the attractive interaction allows one to study the influence of the correlation exponent on the shot noise. Fig.~\ref{fig:variableU} shows the off-diagonal correlator $G_{\uparrow\downarrow}$ of the unpolarized system for interactions $U/t= -1, -2, -4,$ and $-10$ at filling $n=3/4$. According to Ref.~\onlinecite{giamarchi03}, these values of the interaction approximately correspond to Luttinger exponents $K_\rho=$ 1.1, 1.2, 1.3, and 1.4. Note that SS correlations decay with a power-law exponent $1/K_\rho$. For small $U$, the two peaks are very narrow and strongly confined to the area surrounding opposite Fermi points. With increasing $U$, the interaction can scatter particles from well below the Fermi surface, which leads to the spreading out of the base of the peaks. For strong interactions, the Fermi points are barely recognizable, the ridge extending almost uniformly along the whole anti-diagonal. It is thus the width of a peak or dip rather than its height that allows one to make a statement about the strength of the interaction and thus the correlation exponents. Noting that widths and magnitudes are related by a sum rule that ensures vanishing volume under the shot noise~\cite{luscher07}; it is clear that with increasing width, the height of the peaks must decrease, as can be seen in Fig.~\ref{fig:variableU}.

\subsection{Effects of a harmonic trap}
In experiments, the atomic cloud is confined in a harmonic trap of the form
\begin{equation}
H_\text{trap} = V \sum_{l=1}^L \left(l-\frac{L+1}{2}\right)^2 n_i \ .
\end{equation}
with a trapping potential V. Several groups~\cite{feiguin07,tezuka07,batrouni07} have recently analyzed the FFLO pairing in 1D systems 
subject to a confining potential. Compared to the homogeneous case, the FFLO phase does not extend over the whole
chain, but is confined to an interior part of the system: For small densities, an FFLO state is formed in the center of the system,
surrounded by fully polarized wings of unpaired majority particles~\cite{feiguin07,tezuka07}. In contrast, for denser fillings, i.e., 
$n\approx 1$, the fully polarized wings remain, but the FFLO phase is pushed away from the center of the trap and partly replaced
by a polarized metallic phase of freely moving minority particles in a uniform background of majority fermions~\cite{feiguin07}.
In order to detect an FFLO signal in the noise correlations under realistic experimental conditions, it is thus advantageous 
to work in the low-density region. 
To study the effects of a harmonic potential in this regime, we have calculated the 
local densities for systems with $L=64$ sites in the strong-coupling limit $U/t=-10$ at filling $n=1/2$ and polarization $p=1/2$ 
for various depths of the trap, see Figs.~\ref{fig:trap}[row (a)]. Even for shallow confinement [Fig.~\ref{fig:trap}(ai)], the polarized 
wings are already visible. 
It is clear that by pushing some of the majority particles away from the central region of the trap, one
changes the density and the polarization in the FFLO region. It is useful to introduce an effective density $n_\text{eff}=N_\text{eff}/L_\text{eff}$ and polarization $p_\text{eff}=\Delta N_\text{eff}/N_\text{eff}$ at the center of the trap, where the density of the minority spins is greater than some small arbitrary cutoff $\epsilon=10^{-3}$, by defining
\begin{align*}
N_\text{eff} &= \sum_i \theta(n_{i,\downarrow}-\epsilon) \left(n_{i,\uparrow}+n_{i,\downarrow}\right) \ , \\
\Delta N_\text{eff} &= \sum_i  \theta(n_{i,\downarrow}-\epsilon) \left(n_{i,\uparrow}-n_{i,\downarrow}\right) \ ,
\intertext{together with the effective length of the core}
L_\text{eff} &= \sum_i \theta(n_{i,\downarrow}-\epsilon) \ ,
\end{align*}
$\theta$ denoting the usual step function. As illustrated in Figs.~\ref{fig:trap}[row (a)], the effective density 
in the center of the trap increases with increasing confinement, while at the same time the effective polarization is reduced.
The presence of FFLO pairing correlations in the central region of the trap can be deduced from the characteristic signal present in the noise 
correlations $G_{\uparrow\downarrow}$ shown in Figs.~\ref{fig:trap}[row (b)]. In fact, one can convince oneself that the separation between the
two ridges corresponds very accurately to the momentum of the pairs in the middle of the trap $Q_\text{eff}=n_\text{eff} \; p_\text{eff} \; \pi$. 
For a shallow trap, the effect is only subtle, because an increase in density is accompanied by a reduction of the polarization. 
Both the increased density as well as the reduced polarization are unfavorable to the visibility of the FFLO signal in the noise correlations. First, the correlations decay faster than in the homogeneous system, see Fig.~\ref{fig:exponents}(a), and, second,
because the signal is proportional to the number of paired particles, there is a further reduction of the 
FFLO signal due to the smaller effective volume $v_\text{eff}=N_\text{eff}/N$ in which pairing correlations dominate. Despite these adverse effects, 
 the total shot noise shown in Figs.~\ref{fig:trap}[row (c)] clearly illustrates that, for moderate traps, the FFLO signal is still detectable. 

\section{Conclusion}
To conclude, we have derived the correlation exponents of the
1D attractive Hubbard model at finite spin polarizations and have
shown that the FFLO type pairing correlations are dominant for polarizations
$p>0$. Furthermore, we have analyzed the spatial noise correlations
(density-density correlations in momentum space) in view of
experimental realizations with ultra-cold atoms. We have found that
the 1D analog of the inhomogeneous FFLO state leaves a distinct
fingerprint in the shot noise. 

We are grateful to T.~Giamarchi for valuable discussion. This work was supported by the Swiss National Science Foundation.


\begin{thebibliography}{99}

\bibitem{bardeen57}
J.~Bardeen, L.~N. Cooper, and J.~R. Schrieffer, Phys. Rev. {\bf 108}, 1175 (1957).

\bibitem{sarma63}
G.~Sarma, 
J. Phys. Chem. Solids {\bf 24}, 1029 (1963).
\bibitem{fulde64}
P.~Fulde and R.~A. Ferrell, 
Phys. Rev. {\bf 135}, A550 (1964).
\bibitem{larkin64}
A.~I. Larkin and Y.~N. Ovchinnikov, 
Zh. Eksp. Teor. Fiz. {\bf 47}, 1136 (1964) [Sov. Phys. JETP {\bf 20}, 762 (1965)].
\bibitem{muther02}
H.~M\"uther and A.~Sedrakian,
Phys. Rev. Lett. {\bf 88}, 252503 (2002).
\bibitem{liu03}
W.~V. Liu and F.~Wilczek,
Phys. Rev. Lett. {\bf 90}, 047002 (2003).
\bibitem{bedaque03}
P.~F. Bedaque, H.~Caldas, and G.~Rupak, 
Phys. Rev. Lett. {\bf 91}, 247002 (2003).
\bibitem{yi06}
W.~Yi and L.-M.~Duan,
Phys. Rev. Lett. {\bf 97}, 120401 (2006).

\bibitem{radovan03} 
H.~A. Radovan, N.~A. Fortune, T.~P. Murphy, S.~T. Hannahs, E.~C. Palm, S.~W. Tozer, and D.~Hall, 
Nature (London) {\bf 425}, 51 (2003).
\bibitem{bianchi03}
A.~Bianchi, R.~Movshovich, C.~Capan, P.~G. Pagliuso, and J.~L. Sarrao, 
Phys. Rev. Lett. {\bf 91}, 187004 (2003).

\bibitem{casalbuoni04}
R.~Casalbuoni and G.~Nardulli, 
Rev. Mod. Phys. {\bf 76}, 263 (2004).

\bibitem{zwierlein06}
M.~W. Zwierlein, A.~Schirotzek, C.~H. Schunck, and W.~Ketterle, 
Science {\bf 311}, 492 (2006).
\bibitem{partridge06}
G.~B. Partridge, W.~Li, R.~I. Kamar, Y.~Liao, and R.~G. Hulet, 
Science {\bf 311}, 503 (2006).
\bibitem{zwierlein06c}
M.~W. Zwierlein, C.~H. Schunck, A.~Schirotzek, and W.~Ketterle,
Nature {\bf 442}, 54 (2006).
\bibitem{shin06}
Y.~Shin, M.~W. Zwierlein, C.~H. Schunck, A.~Schirotzek, and W.~Ketterle,
Phys. Rev. Lett. {\bf 97}, 030401 (2006).
\bibitem{schnuck07}
C.~H. Schnuck, Y.~Shin, A.~Schirotzek, M.~W. Zwierlein, and W.~Ketterle,
Science {\bf 316}, 867 (2007).

\bibitem{mizushima05}
T. Mizushima, K. Machida, and M. Ichioka, 
Phys. Rev. Lett. {\bf  94}, 060404 (2005).
\bibitem{orso07}
G.~Orso,
Phys. Rev. Lett. {\bf 98}, 070402 (2007).
\bibitem{hu07}
H.~Hu, X.-J. Liu, and P.~D. Drummond,
Phys. Rev. Lett. {\bf 98}, 070403 (2007).
\bibitem{parish07}
M.~M. Parish, S.~K. Baur, E.~J. Mueller, and D.~A. Huse,
Phys. Rev. Lett. {\bf 99}, 250403 (2007).

\bibitem{moritz05}
H.~Moritz, T.~St\"oferle, K.~G\"unter, M.~K\"ohl, and T.~Esslinger,
Phys. Rev. Lett. {\bf 94}, 210401 (2005).

\bibitem{chandrasekhar62}
B.~S. Chandrasekhar,
Appl. Phys. Lett. {\bf 1}, 7 (1962).
\bibitem{clogston62}
A.~M. Clogston, 
Phys. Rev. Lett. {\bf 9}, 266 (1962).

\bibitem{altman04}
E.~Altman, E.~Demler, and M.~D. Lukin,
Phys. Rev. A {\bf 70}, 013603 (2004). 
\bibitem{yang05}
K.~Yang,
Phys. Rev. Lett. {\bf 95}, 218903 (2005).

\bibitem{singh91}
R.~R.~P. Singh and R.~T. Scalettar,
Phys. Rev. Lett. {\bf 66}, 3203 (1991).

\bibitem{yang01}
K. Yang,
Phys. Rev. B {\bf 63}, 140511(R) (2001).

\bibitem{emery76}
V.~J. Emery, 
Phys. Rev. B {\bf 14}, 2989 (1976).
\bibitem{moreo07}
A.~Moreo and D.~J. Scalapino,
Phys. Rev. Lett. {\bf 98}, 216402 (2007).
\bibitem{frahm}
H.~Frahm and V.~E. Korepin,
Phys. Rev. B {\bf 42}, 10553 (1990); ibid.
{\bf 43}, 5653 (1991).
\bibitem{giamarchi03}
T.~Giamarchi,
{\it Quantum Physics in One Dimension}, 
Oxford University Press (2003).
\bibitem{frahm08}
H.~Frahm and T.~Vekua,
J. Stat. Mech.: Theory Exp. ({\bf 2008}) P01007. 
\bibitem{rizzi07}
M.~Rizzi, M.~Polini, M.~A. Cazalilla, M.~R. Bakhtiari, M.~P. Tosi, and R.~Fazio,
Phys. Rev. B {\bf 77}, 245105 (2008).
\bibitem{zhao08}
E.~Zhao and W.~V. Liu,
arXiv:0804.4461v1 (unpublished).

\bibitem{white92}
S.~R.White,
Phys. Rev. Lett. {\bf 69}, 2863 (1992);
Phys. Rev. B {\bf 48}, 10345 (1993).
\bibitem{schollwoeck05}
U.~Schollw\"ock,
Rev. Mod. Phys. {\bf 77}, 259 (2005).

\bibitem{luscher07}
A.~L\"uscher, A.~M. L\"auchli, and R.~M. Noack,
Phys. Rev. A {\bf 76}, 043614 (2007).
\bibitem{mathey05}
L.~Mathey, E.~Altman, and A.~Vishwanath,
Phys. Rev. Lett. {\bf 100}, 240401 (2008).

\bibitem{batrouni07}
G.~G. Batrouni, M.~H. Huntley, V.~G. Rousseau, and R.~T. Scalettar,
Phys. Rev. Lett. {\bf 100}, 116405 (2008).
\bibitem{feiguin07}
A.~E. Feiguin and F.~Heidrich-Meisner,
Phys. Rev. B {\bf 76}, 220508 (2007).
\bibitem{tezuka07}
M.~Tezuka and M.~Ueda,
Phys. Rev. Lett. {\bf 100}, 110403 (2008).

\end{thebibliography}
\end{document}